\documentclass[twocolumn,showpacs,amsmath,amssymb]{revtex4}

\usepackage{graphicx}
\usepackage{bm}

\newcommand{\dif}{\,\mathrm{d}}
\renewcommand{\Im}{\,\mathrm{Im}}
\newcommand{\ldirac}[1]{\langle #1 |}
\newcommand{\rdirac}[1]{|#1\rangle}

\begin{document}

\title{
The van der Waals energy of atomic systems near
absorbing and dispersing bodies}

\author{Stefan Yoshi Buhmann}

\author{Ho Trung Dung}
\altaffiliation[Also at]{
Institute of Physics, National Center
for Sciences and Technology, 1 Mac Dinh Chi Street,
District 1, Ho Chi Minh city, Vietnam.}

\author{Dirk-Gunnar Welsch}

\affiliation{Theoretisch-Physikalisches Institut,
Friedrich-Schiller-Universit\"{a}t Jena,
Max-Wien-Platz 1, 07743 Jena, Germany}

\date{\today}

\begin{abstract}
Within the frame of macroscopic quantum electrodynamics
in causal media, the van der Waals interaction between
an atomic system and an arbitrary arrangement of dispersing and
absorbing dielectric bodies including metals is
studied. It is shown that the minimal-coupling scheme
and the multipolar-coupling scheme lead to
essentially the same formula for the van der Waals potential.
As an application, the vdW potential of an atom in the presence of a
sphere is derived.
Closed expressions for the long-distance (retardation)
and short-distance (non-retardation) limits are given,
and the effect of material absorption is discussed.
\end{abstract}

\pacs{
12.20.-m, 
42.50.Vk, 
42.50.Nn 
}

\maketitle

\section{Introduction}
\label{Sec:Intro}

The involvement of van der Waals (vdW) forces
in a variety of physicochemical processes and promising potential
applications (such as the construction of atomic-force microscopes
\cite{Binnig86} or reflective atom-optical elements \cite{Shimizu02})
have created the need for a very detailed understanding and
controlling them.
Casimir and Polder \cite{CasimirPolder} were the first to study the
vdW interaction within the frame of rigorous
quantum electrodynamics (QED). Investigating the interaction
between an atom and a perfectly reflecting
semi-infinite (planar) body, they found that in the
short-distance (non-retardation) limit the interaction potential $U(z)$
behaves like $z^{-3}$ ($z$, distance between the atom and
the interface), whereas in the long-distance (retardation) limit it
behaves like $z^{-4}$. In their theory,
Casimir and Polder quantized the (transverse) vector potential of the
electromagnetic field (outside the perfectly reflecting body)
in terms of normal modes and coupled them to the atom according
to the minimal-coupling Hamiltonian, with its
Coulomb part being determined by means of
the method of image charges. They then
calculated the (lowest-order) change of the ground-state energy
of the system arising from this coupling, which is a function of the
atomic position and thus plays the role of the potential
energy that determines the force acting on the atom.

In the earlier experiments \cite{Raskin69}, which studied
the deflection  of thermal atomic beams by conducting surfaces,
the observed signal was extremely low. Nevertheless, qualitative
trends in agreement with the $z^{-3}$ law were observed.
Only recent progress in experimental techniques has rendered
it possible to detect vdW forces with sufficiently high precision
\cite{AndersonHarocheHinds,GrisentiSchoellkopfToennies,%
Shimizu01,LandraginCourtoisLabeyrie,SandoghdarSukenikHinds}.
In particular, by using atomic passage between two parallel plates,
the $z^{-4}$ retarded potential could be verified
\cite{AndersonHarocheHinds}.
Other methods for measuring vdW forces have been based on
transmission grating diffraction of molecular beams
\cite{GrisentiSchoellkopfToennies},
atomic quantum reflection \cite{Shimizu01},
evanescent-wave atomic mirror techniques
\cite{LandraginCourtoisLabeyrie}, and (indirect)
measurements via spectroscopic means
\cite{SandoghdarSukenikHinds}. Proposals have
been made on improvements of monitoring the vdW interaction
by using atomic interferometry
\cite{GorlickiFeronLorentDucloy}.

Since the appearance of Casimir's and Polder's pioneering article
in 1948 there has been a large body of work on the
vdW interaction (see, e.g.,
\cite{Dzyaloshinskii61,Langbein74,Mahanty76,Hinds91,Milonni94}
and references therein). Roughly speaking, there have been
two routes to treat the problem. In the first, which closely
follows the ideas of Casimir and Polder,
explicit field quantization is performed by applying
standard concepts of QED, such as normal-mode techniques
\cite{Bullough70,Renne,Milloni,Tikochinski,Zhou,Bostrom00,MarvinToigo,Wu}.
In particular, extensions of the theory to one \cite{Milloni,Renne}
and two semi-infinite dielectric walls \cite{Tikochinski,Zhou},
thin metallic films \cite{Bostrom00}, and cylindrical and spherical
dielectric bodies \cite{MarvinToigo} have been given,
and the problem of force fluctuations on short time scales
has been studied \cite{Wu}.
The calculations have typically been based on macroscopic QED,
by applying normal-mode decomposition and including
in it the bodies by the well-known conditions
of continuity at the surfaces of discontinuity.
Since in such an approach the dependence on frequency
of the response to the field of the bodies cannot
be properly taken into account, material dispersion and
absorption are ignored. The problem does of course not
occur in microscopic QED, where the bodies are treated
on a microscopical level by adopting, e.g.,
harmonic-oscillator models (see, e.g., \cite{Renne}).
Apart from the fact that the calculations are rather involved,
the results are model-dependent.

To overcome the difficulties mentioned,
in the second route, the calculations are based on linear
response theory, without (explicitly) quantizing
the electromagnetic field
\cite{McLachlan, Argawal2,WylieSipe,Girard,GirardGirardet,Fichet,
GirardMaghezzi,Boustimi}. All the relevant entities
are expressed in terms of correlation functions which in turn are
related, via the fluctuation-dissipation theorem,
to response functions. The method
has been employed to investigate the vdW energy of an atom near
a semi-infinite (planar) body made of dielectric
\cite{McLachlan,Argawal2}, metallic \cite{WylieSipe,Girard},
ionic-crystallic \cite{GirardGirardet}, and birefringent
material \cite{Fichet}. Further, the problem of an atom  near a
a dielectric plate \cite{WylieSipe}, a metallic sphere
\cite{GirardMaghezzi}, and a nanowire \cite {Boustimi}
has been considered.
Since the method borrows from equilibrium statistical
mechanics, its applicability is restricted to the vacuum and
thermal quantum states.

In this article, we give a unified QED approach to the vdW
interaction between an atomic system (such as an atom or
a molecule) and dispersing and absorbing dielectric bodies
including metals. Starting from the quantized version of
the macroscopic Maxwell field, with the medium being
described in terms of a spatially varying,
Kramers-Kronig consistent (complex) permittivity, we
derive an expression for the vdW potential that applies to
arbitrary body configurations. The formalism can be regarded as
being a generalization of the normal-mode formalism
of macroscopic QED, so that it can be applied to other
than the vacuum and thermal states, and it also allows
extensions like the inclusion of dynamical interactions
between the atomic system and the medium-assisted
electromagnetic field. Roughly speaking,
the mode expansion is replaced with some source-quantity
representation in terms of the Green tensor of the
macroscopic Maxwell equations, in which material
dispersion and absorption are automatically included.
Further, we close the gap between the minimal-coupling
scheme and the multipolar-coupling scheme
by showing that both approaches lead to equivalent results.

To give an application, we consider the vdW
potential of an atom in the vicinity of a
dispersing and absorbing microsphere.
Microspheres may be interesting candidates
for QED experiments (see, e.g., \cite{Buck02,Ho01} and
references therein) with atomic beams.
The atom-surface distances should be
adjusted so that the atoms fly close enough
to the surface to facilitate a strong coupling with the microsphere
resonances, but not get adsorbed on the microsphere surface
because of the attractive vdW force.

The article is organized as follows. In Section \ref{Sec:Basic_eqs}
the formalism is outlined and an expression for the vdW potential
is derived. In Section \ref{Sec:Appl} the formalism is applied
to an atom near a sphere. A summary and some conclusions are given
in Section \ref{Sec:Concl}.


\section{The van der Waals energy}
\label{Sec:Basic_eqs}

\subsection{The quantization scheme}
\label{Subsec:mincH}

Let us consider an atomic system (such as an atom or a molecule)
interacting with the quantized electromagnetic field in
the presence of macroscopic, dispersing and absorbing dielectric
bodies. In the nonrelativistic limit, the minimal-coupling
Hamiltonian in Coulomb gauge reads \cite{Knoll01,Scheel99}
\begin{eqnarray}
\label{E3}
      \hat{H}&=&
      \int_0^{\infty}\dif\omega\,\hbar\omega
      \int\dif^3r\,\hat{\bf f}^{\dagger}({\bf r},\omega)
      \hat{\bf f}({\bf r},\omega)
\nonumber\\
      &&+\sum_{\alpha}\frac{1}{2 m_{\alpha}}
      \left[\hat{\bf p}_{\alpha}
      -q_{\alpha}\hat{\bf A}({{\bf r}_{\rm A}}+\hat{\bf r}_{\alpha})\right]^2
\nonumber\\
      &&
      + {\textstyle\frac{1}{2}}\int\dif^3r\,\hat{\rho}_A({\bf r})
      \hat{\varphi}_{\rm A}({\bf r})
      +\int\dif^3r\,\hat{\rho}_A({\bf r})
      \hat{\varphi}_{\rm M}({\bf r}),
      \quad
\end{eqnarray}
where $m_\alpha$ and $q_\alpha$ are respectively the masses and
charges of the particles constituting the atomic system,
while $\hat{\bf r}_\alpha$ and $\hat{\bf p}_{\alpha}$ are
respectively their coordinates (relative to the center of mass
${{\bf r}_{\rm A}}$) and canonically conjugated momenta. The first
term in the Hamiltonian describes the combined system
of the electromagnetic field plus the macroscopic bodies
(including dissipative systems) in terms
of bosonic vector fields $\hat{\bf f}({\bf r},\omega)$, which
satisfy the commutation relations
\begin{equation}
\label{E4}
      \left[\hat{f}_i({\bf r},\omega),
      \hat{f}_j^{\dagger}({\bf r'},\omega')\right]
      =\delta_{ij}\delta({\bf r}-{\bf r}')\delta(\omega-\omega'),
\end{equation}
\begin{equation}
\label{E5}
      \left[\hat{f}_i({\bf r},\omega),
      \hat{f}_j({\bf r'},\omega')\right]
      =\left[\hat{f}_i^{\dagger}({\bf r},\omega),
      \hat{f}_j^{\dagger}({\bf r'},\omega')\right]
      =0.
\end{equation}
The second term is the kinetic energy of the charged particles,
while the third term describes their mutual Coulomb
interaction, where
\begin{equation}
\label{E6}
\hat{\varphi}_{\rm A}({\bf r})
= \frac{1}{4\pi\varepsilon_0} \int \dif^3{r}'\,
\frac{\hat{\rho}_{\rm A}({\bf r}')}{|{\bf r}-{\bf r}'|}
\end{equation}
and
\begin{equation}
\label{E7}
      \hat{\rho}_{\rm A}({\bf r})
      =\sum_{\alpha}q_{\alpha}\delta[{\bf r}
      -({{\bf r}_{\rm A}}+\hat{\bf r}_{\alpha})]
\end{equation}
are respectively the scalar potential and the charge density
of the atomic system. The last term accounts for the
Coulomb interaction of the particles with the medium.
The vector potential $\hat{\bf A}({\bf r})$ and the scalar
potential $\hat{\varphi}_{\rm M}({\bf r})$ of the medium-assisted
electromagnetic field are given by
\begin{eqnarray}
\label{E8}
      \hat{\bf A}({\bf r}) =
      \int_0^\infty {\rm d} \omega \, (i\omega)^{-1}
      \underline{\hat{\bf E}}{^\perp}({\bf r},\omega)
      + {\rm H.c.},
      \end{eqnarray}
\begin{eqnarray}
\label{E9}
      -\bm{\nabla} \hat{\varphi}_{\rm M}({\bf r})
      = \int_0^\infty {\rm d} \omega \,
      \underline{\hat{\bf E}}{^\parallel}({\bf r},\omega)
      + {\rm H.c.},
\end{eqnarray}
where the symbols $\perp$ and $\parallel$ are used to distinguish
transverse and longitudinal vector fields, respectively.
In particular, $\hat{\underline{\bf E}}{^{\perp}}({\bf r},\omega)$
and $\hat{\underline{\bf E}}{^{\parallel}}({\bf r},\omega)$ read
\begin{equation}
\label{E10}
      \hat{\underline{\bf E}}{^{\perp(\parallel)}}({\bf r},\omega)
      = \int {\rm d}^3 r' \, \mbox{\boldmath $\delta$}
      ^{\perp(\parallel)}({\bf r}-{\bf r}')
      {} \hat{\underline{\bf E}}({\bf r}',\omega),
\end{equation}
where $\mbox{\boldmath $\delta$}^\perp({\bf r})$ and
$\mbox{\boldmath $\delta$}^\parallel({\bf r})$ are the transverse
and longitudinal dyadic $\delta$-functions, respectively,
and the medium-assisted electric field in the
$\omega$-domain, $\hat{\underline{\bf E}}({\bf r},\omega)$,
is expressed in terms of the basic variables
$\hat{\bf f}({\bf r},\omega)$ as
\begin{equation}
\label{E11}
\begin{split}
\underline{\hat{\bf E}}({\bf r},\omega)
      = i \sqrt{\frac{\hbar}{\pi\varepsilon_0}}\,\frac{\omega^2}{c^2}
        \int & {\rm d}^3r'\,\sqrt{{\rm Im}\,\varepsilon({\bf r}',\omega)}
\\[1ex]&\times\;
      \bm{G}({\bf r},{\bf r}',\omega)
      {}\hat{\bf f}({\bf r}',\omega)
\end{split}
\end{equation}
\mbox{($c$ $\!=$ $1/\sqrt{\varepsilon_0\mu_0}$)}.
Here, $\bm{G}({\bf r},{\bf r}',\omega)$ is the classical Green tensor
and ${\rm Im}\,\varepsilon({\bf r},\omega)$ is the imaginary part
of the complex, space- and frequency-dependent
(relative) permittivity
\begin{equation}
\label{E12}
\varepsilon({\bf r},\omega)
= {\rm Re}\,\varepsilon({\bf r},\omega)
+ i{\rm Im}\,\varepsilon({\bf r},\omega).
\end{equation}
The Green tensor, which obeys the inhomogeneous partial differential equation
\begin{equation}
\label{E19}
      \left[
      \bm{\nabla}\times\bm{\nabla}\times
      -\frac{\omega^2}{c^2}\,\varepsilon({\bf r},\omega)
      \right]
      \bm{G}({\bf r},{\bf r}',\omega)
      = \bm{\delta}({\bf r}-{\bf r}')
      \end{equation}
together with the boundary condition at infinity, has the following general properties (see, e.g., \cite{Knoll01}):
\begin{equation}
\label{E20}
     \bm{G}^{\ast}({\bf r},{\bf r}',\omega)
     =\bm{G}({\bf r},{\bf r}',-\omega^{\ast}),
\end{equation}
\begin{equation}
\label{E21}
     G_{ij}({\bf r},{\bf r}',\omega)
     =G_{ji}({\bf r}',{\bf r},\omega),
\end{equation}
\begin{equation}
\label{E22}
     \frac{\omega^2}{c^2}\!
     \int\!\dif^3{s}\,\varepsilon_{\rm I}({\bf s},\omega)
     \bm{G}({\bf r},{\bf s},\omega)\bm{G}^{\ast}({\bf s},{\bf r}',\omega)
     =\Im \bm{G}({\bf r},{\bf r}',\omega).
\end{equation}

In this way, all electromagnetic-field quantities can be expressed
in terms of the fundamental fields $\hat{\bf f}({\bf r},\omega)$.
In particular, the operator for the electric field reads
\begin{eqnarray}
\label{E13}
 \hat{\bf E}({\bf r})
 &=& -\frac{1}{i\hbar}\left[\hat{\bf A}({\bf r}),\hat{H}\right]
 - \bm{\nabla}\hat{\varphi}_{\rm M}({\bf r})
 - \bm{\nabla}\hat{\varphi}_{\rm A}({\bf r})
\nonumber\\
  &=& \hat{\bf E}_{\rm M}({\bf r})
   + \sum_\alpha \frac{q_\alpha[{\bf r}-({\bf r}_{\rm A}+\hat{\bf r}_\alpha)]}
   {4\pi\varepsilon_0|{\bf r}-({\bf r}_{\rm A}+\hat{\bf r}_\alpha)|^3}\,,
\end{eqnarray}
where
\begin{equation}
\label{E14}
\hat{\bf E}_{\rm M}({\bf r})
= \int_0^{\infty}\dif\omega\,
\underline{\hat{\bf E}}({\bf r},\omega) + {\rm H.c.},
\end{equation}
and the polarization field associated with the dielectric medium is given by
\begin{equation}
\label{E15}
 \hat{\bf P}_{\rm M}({\bf r})
 = \int_0^{\infty}\dif\omega\,\underline{\hat{\bf P}}
 ({\bf r},\omega)+{\rm H.c.},
\end{equation}
where
\begin{equation}
\label{E16}
\underline{\hat{\bf P}}({\bf r},\omega)
= \chi({\bf r},\omega)\varepsilon_0
\underline{\hat{\bf E}}({\bf r},\omega)
+ \hat{\bf P}_{\rm N}({\bf r},\omega).
\end{equation}
Here,
\begin{equation}
\label{E17}
\chi({\bf r},\omega) = \varepsilon({\bf r},\omega)-1
\end{equation}
is the dielectric susceptibility and
\begin{equation}
\label{E18}
 \hat{\bf P}_{\rm N}({\bf r},\omega)
 =i\sqrt{\frac{\hbar\varepsilon_0}{\pi}\,
 {\rm Im}\,\varepsilon({\bf r},\omega)}
 \,\hat{\bf f}({\bf r},\omega)
\end{equation}
is the so-called noise polarization.

For the following it is convenient to decompose the
Hamiltonian (\ref{E3}) as
\begin{equation}
\label{E23}
     \hat{H} =  \hat{H}_{\rm F} + \hat{H}_{\rm A} + \hat{H}_{\rm AF},
\end{equation}
where
\begin{equation}
\label{E24}
      \hat{H}_{\rm F} \equiv \int_0^{\infty}\dif\omega\,\hbar\omega
      \int\dif^3 {r}\,
      \hat{\bf f}^{\dagger}({\bf r},\omega)
      \hat{\bf f}({\bf r},\omega),
\end{equation}
\begin{eqnarray}
\label{E25}
      \hat{H}_{\rm A}
      \hspace{-1ex}&\equiv&\hspace{-1ex}
      \sum_{\alpha}
      \frac{\hat{{\bf p}}_{\alpha}^2}{2m_{\alpha}}
      +\textstyle{\frac{1}{2}}\int {\rm d}^3 r\,
      \hat{\rho}_{\rm A}({\bf r})\hat{\varphi}_{\rm A}({\bf r})
\nonumber\\[1ex]
      \hspace{-1ex}&=&\hspace{-1ex}
      \sum_{\alpha}
      \frac{\hat{{\bf p}}_{\alpha}^2}{2m_{\alpha}}
      +\sum_{\alpha<\beta} \frac{q_{\alpha}q_{\beta}}
      {4\pi\varepsilon_0\left|\hat{{\bf r}}_{\alpha}
      -\hat{{\bf r}}_{\beta}\right|}\,,
\end{eqnarray}
\begin{eqnarray}
\label{E26}
      \hat{H}_{\rm AF} &\equiv&
      \int\dif^3{r}\,\hat{\rho}_A({\bf r})
      \hat{\varphi}_{\rm M}({\bf r})-\sum_{\alpha}
      \frac{q_{\alpha}}{m_{\alpha}}\,
      \hat{\bf A}({{\bf r}_{\rm A}}+\hat{\bf r}_{\alpha})
\nonumber\\&&\times
      \left[\hat{\bf p}_{\alpha}
      -\textstyle{\frac{1}{2}}q_{\alpha}
      \hat{\bf A}({{\bf r}_{\rm A}}+\hat{\bf r}_{\alpha})\right].
      \quad
\end{eqnarray}
Obviously, $\hat{H}_{\rm F}$ is the Hamiltonian of the medium-assisted
electromagnetic field, $\hat{H}_{\rm A}$ is the Hamiltonian of the
atomic system with eigenstates $\rdirac{n}$ and eigenvalues $E_n$
according to
\begin{equation}
\label{E27}
       \hat{H}_{\rm A}\rdirac{n}=E_n\rdirac{n},
\end{equation}
and $\hat{H}_{\rm AF}$ is the interaction energy between them.
For a neutral atomic system in the electric-dipole approximation, the latter simplifies to
\begin{equation}
\label{E28}
      \hat{H}_{\rm AF} = \hat{H}_{\rm AF}^{\rm (I)}
      + \hat{H}_{\rm AF}^{\rm (II)}\,,
\end{equation}
\begin{equation}
\label{E29}
      \hat{H}_{\rm AF}^{\rm (I)} \equiv
      -\sum_{\alpha}\frac{q_{\alpha}}{m_{\alpha}}
      \,\hat{\bf p}_{\alpha}\hat{\bf A}({{\bf r}_{\rm A}})
      +\hat{\bf d}\bm{\nabla}\hat{\varphi}_{\rm M}({{\bf r}_{\rm A}}),
\end{equation}
\begin{equation}
\label{E30}
 \hat{H}_{\rm AF}^{\rm (II)} \equiv
      \sum_{\alpha}\frac{q_{\alpha}^2}{2m_{\alpha}}
      \,\hat{\bf A}^2({{\bf r}_{\rm A}}),
\end{equation}
where $\hat{\bf d}$ is the electric dipole operator of
the atomic system
\begin{equation}
\label{E31}
      \hat{\bf d}= \sum_{\alpha}q_{\alpha}\hat{\bf r}_{\alpha}.
\end{equation}

\subsection{The vdW energy in the minimal-coupling scheme}
\label{Subsec:vdW_mincf}

Following the original line of Casimir and Polder
\cite{CasimirPolder}, we calculate the vdW energy of
an atomic system as the
position-dependent part of the leading-order correction to the
unperturbed ground state energy due to the perturbation
according to Eq.~(\ref{E28}) [together with Eqs.~(\ref{E29})
and (\ref{E30})].  
Since the diagonal matrix elements of $\hat{H}_{\rm AF}^{\rm (I)}$,
Eq.~(\ref{E29}), are zero, the non-vanishing contribution to
the energy correction from first-order perturbation theory
is due to $\hat{H}_{\rm AF}^{\rm (II)}$, Eq.~(\ref{E30}).
Inspection of Eq.~(\ref{E30}) shows that this contribution
is of linear order in the fine structure constant
\mbox{$\alpha$ $\!=$ $\!e^2/(2\varepsilon_0\hbar c)$}. We have thus to
include that contribution from second-order perturbation theory
which is of the same order in $\alpha$.
Therefore, we apply first-order perturbation theory
for $\hat{H}_{\rm AF}^{\rm (II)}$ and second-order perturbation
theory for $\hat{H}_{\rm AF}^{\rm (I)}$.
The energy correction to the ground state thus reads
\begin{equation}
\label{E32}
      \Delta E \simeq \Delta_1 E + \Delta_2 E,
\end{equation}
where
\begin{eqnarray}
\label{E33}
      \Delta_1 E
      &\equiv&\ldirac{0}\ldirac{\{0\}}
      \hat{H}_{AF}^{\rm (II)}\rdirac{\{0\}}\rdirac{0}
\nonumber \\[1ex]
      &=&\ldirac{0}\ldirac{\{0\}}
      \sum_{\alpha}\frac{q_{\alpha}^2}{2m_{\alpha}}
      \,\hat{\bf A}^2({{\bf r}_{\rm A}})
      \rdirac{\{0\}}\rdirac{0}
\end{eqnarray}
and
%
%
\begin{eqnarray}
\label{E34}
\lefteqn{
      \Delta_2 E
      \equiv
      \sum_n\int_0^{\infty}\!\!\dif\omega\!\int\!\dif^3{r}
      \;\frac{|\ldirac{0}\ldirac{\{0\}}
             \hat{H}_{\rm AF}^{\rm (I)}
             \rdirac{\{{\bf 1}({\bf r},\omega)\}}\rdirac{n}|^2}
      {E_0-(E_n+\hbar\omega)}
}      
\nonumber\\[1ex]&&
      =\frac{1}{\hbar}\sum_n\int_0^{\infty}\!\!
      \frac{\dif\omega}{\omega_n+\omega} \int\!\dif^3{r}
      \,\Bigl
      |\ldirac{0}\ldirac{\{0\}}\sum_{\alpha}\frac{q_{\alpha}}{m_{\alpha}}
      \hat{\bf p}_{\alpha}\hat{\bf A}({{\bf r}_{\rm A}})
\nonumber\\[1ex]&&\hspace{10ex}
      -\,\hat{\bf d}\bm{\nabla}\hat{\varphi}_{\rm M}({{\bf r}_{\rm A}})
      \rdirac{\{{\bf 1}({\bf r},\omega)\}}\rdirac{n}
     \Bigr
     |^2.
\end{eqnarray}
Here, $|0\rangle$ and $|\{0\}\rangle$ are respectively the ground
state of the atomic system and the ground state of the
medium-assisted electromagnetic field, and
\begin{equation}
\label{E35}
|\{{\bf 1}({\bf r},\omega)\}\rangle
\equiv \hat{\bf f}{^{\dagger}}({\bf r},\omega)|\{0\}\rangle
\end{equation}
denotes single-quantum Fock states of the fundamental fields. Further,
\begin{equation}
\label{E36}
      \omega_n = (E_n-E_0)/\hbar
\end{equation}
are the transition frequencies between the excited atomic states
and the ground state. Note that due to the linear dependence of
the vector potential and the gradient of the scalar potential on
$\hat{\bf f}({\bf r},\omega)$ [and $\hat{\bf f}^\dagger({\bf r},
\omega)$], the nonvanishing matrix elements of
$\hat{H}_{\rm AF}^{\rm (I)}$ in the Fock-state basis
(which is defined as the set of eigenvectors of $\hat{H}_{\rm F}$)
are those between Fock states that just differ in one quantum.

Expressing in Eq.~(\ref{E33}) $\hat{\bf A}({{\bf r}_{\rm A}})$
in terms of $\hat{\bf f}({\bf r})$ [and
$\hat{\bf f}^\dagger({\bf r})$], on using Eqs.~(\ref{E8}),
(\ref{E10}), and (\ref{E11}), recalling Eq.~(\ref{E35}), and
exploiting the integral relation (\ref{E22}) for the contraction
of two Green tensors, we derive after some lengthy but
straightforward calculation
\begin{equation}
\label{E37}
     \Delta_1 E =
     \frac{\hbar\mu_0}{\pi}\sum_{\alpha}
     \frac{q_{\alpha}^2}{2m_{\alpha}}
     \int_0^{\infty}\dif\omega
     \Im ^\perp G^\perp_{ii}({{\bf r}_{\rm A}},{{\bf r}_{\rm A}},\omega),
\end{equation}
where
\begin{equation}
\label{E38}
\begin{split}
     &^{\perp(\parallel)}\bm{G}^{\perp(\parallel)}({\bf r},{\bf r}',\omega)
     \\&\quad
     \equiv \!\int\!\dif^3{s}\!\int\!\dif^3{s}'\,
     \bm{\delta}^{\perp(\parallel)}({\bf r}-{\bf s})
     \bm{G}({\bf s},{\bf s}',\omega)
     \bm{\delta}^{\perp(\parallel)}({\bf s'}-{\bf r}').
     \quad
\end{split}
\end{equation}
In Eq. (\ref{E37}) and below, summation over repeated
vector indices is understood. Using the sum rule
\begin{equation}
\label{E39}
       \sum_{\alpha}\frac{q_{\alpha}^2}{2m_{\alpha}}\delta_{ij}
       = \frac{1}{2\hbar}\sum_n\omega_n
       (d_{0n,i} d_{n0,j} + d_{0n,j} d_{n0,i})
\end{equation}
(for a proof, see Appendix \ref{Sec:sumrule}), where
\begin{equation}
\label{E42}
    d_{0n,i} \equiv \langle 0| \hat{d}_i |n\rangle,
\end{equation}
denote the matrix elements of the electric-dipole operator (\ref{E31}),
we may equivalently represent Eq.~(\ref{E37}) in the form of
\begin{eqnarray}
\label{E40}
\lefteqn{    
     \Delta_1 E =
     \frac{\hbar\mu_0}{\pi}\sum_{\alpha}
     \frac{q_{\alpha}^2}{2m_{\alpha}} \delta_{ij}
     \int_0^{\infty}\dif\omega
     \,\Im ^\perp G^\perp_{ij}({{\bf r}_{\rm A}},{{\bf r}_{\rm A}},\omega)
}
\nonumber\\[1ex]&&  
     =\frac{\mu_0}{\pi}
     \sum_n \int_0^{\infty}\dif\omega\, \omega_n
     {\bf d}_{0n}
     \Im ^{\perp}\bm{G}^{\perp}({{\bf r}_{\rm A}},{{\bf r}_{\rm A}},\omega)
     {\bf d}_{n0}.
     \qquad
\end{eqnarray}
In a similar fashion, we find that Eq.~(\ref{E34}) leads to
\begin{equation}
\label{E41}
\begin{split}
     &\Delta_2 E
     =
     - \frac{\mu_0}{\pi}\sum_n
     \int_0^{\infty}\dif\omega\,
     \biggl[
     \frac{\omega_n^2}{\omega_n+\omega}
     \int\dif^3{s}
     \\[1ex]
     &\times\int\dif^3{s}'
     \,{\bf d}_{0n} \bm{\mu}({\bf s},\omega)
     {\rm Im}\,\bm{G} ({\bf s},{\bf s}',\omega)  \bm{\mu}({\bf s}',\omega)
     {\bf d}_{n0}\biggr],
\end{split}
\end{equation}
where, on using the relation (\ref{A3}), the matrix elements
of the electric-dipole moment (\ref{E42}) have been
introduced and the abbreviating notation
\begin{equation}
\label{E43}
      \bm{\mu}({\bf r},\omega) \equiv
      \bm{\delta}^\perp({\bf r}-{{\bf r}_{\rm A}})
      - \frac{\omega}{\omega_n}\,
      \bm{\delta}^\parallel({\bf r}-{{\bf r}_{\rm A}})
\end{equation}
has been used.

We substitute Eqs.~(\ref{E40}) and (\ref{E41}) into Eq.~(\ref{E32}),
and obtain, on recalling Eq.~(\ref{E43}),
\begin{eqnarray}
\label{E44}
\lefteqn{
     \Delta E =
     -\frac{\mu_0}{\pi}
     \sum_n \int_0^{\infty} \frac{\dif\omega}{\omega_n+\omega}
     \,{\bf d}_{0n}\Bigl\{\omega^2
     \Im ^\parallel\bm{G}^\parallel({{\bf r}_{\rm A}},{{\bf r}_{\rm A}},\omega)
}\quad
\nonumber\\[1ex]&&\hspace{3ex}
     -\,\omega_n\omega\bigl[
     \Im ^\perp\bm{G}^\perp({{\bf r}_{\rm A}},{{\bf r}_{\rm A}},\omega)
     +\Im ^\perp\bm{G}^\parallel({{\bf r}_{\rm A}},{{\bf r}_{\rm A}},\omega)
\nonumber\\[1ex]&&\hspace{3ex}
     +\,\Im ^\parallel\bm{G}^\perp({{\bf r}_{\rm A}},
     {{\bf r}_{\rm A}},\omega) \bigr]
     \Bigr\} {\bf d}_{n0}.
\end{eqnarray}
Let ${\cal R}$ be the (small) region of space where the atom
is situated and the permittivity can be regarded as
being effectively not varying with space, i.e.,
\mbox{$\varepsilon({\bf r},\omega)$ $\!\simeq$ $\!\varepsilon({{\bf
r}_{\rm A}},\omega)$} if \mbox{${\bf r}$ $\!\in$ $\!\mathcal{R}$}.
For \mbox{${\bf r},{\bf r}'$ $\!\in$
$\!\mathcal{R}$}, the Green tensor can then be given in the form
\begin{equation}
\label{E45}
    \bm{G}({\bf r},{\bf r}',\omega)
    =\bm{G}^{(0)}({\bf r},{\bf r}',\omega)
    +\bm{G}^{(1)}({\bf r},{\bf r}',\omega),
\end{equation}
with $\bm{G}^{(0)}({\bf r},{\bf r}',\omega)$ denoting the
(translationally invariant) bulk Green tensor that corresponds to
$\varepsilon({{\bf r}_{\rm A}},\omega)$ and $\bm{G}^{(1)}({\bf r},
{\bf r}',\omega)$ being the scattering Green tensor that
accounts for the spatial variation of the permittivity.
In practice, the atom is typically situated in a free-space
region, so that $\bm{G}^{(0)}$ is simply the vacuum Green tensor.
According to the decomposition of the Green tensor in
Eq.~(\ref{E45}), the energy correction $\Delta E$ given
by Eq.~(\ref{E44}) consists of two terms,
\begin{equation}
\label{E46}
\Delta E = \Delta E^{(0)} + \Delta E^{(1)}({{\bf r}_{\rm A}}),
\end{equation}
where the ${{\bf r}_{\rm A}}$-independent term $\Delta E^{(0)}$,
which is related to the bulk Green tensor, gives rise
to the (vacuum) Lamb shift, whereas the ${{\bf r}_{\rm A}}$-dependent
term $\Delta E^{(1)}({{\bf r}_{\rm A}})$, which is related to
the scattering Green tensor, is just the vdW energy sought:
\begin{eqnarray}
\label{E47}
     &&U({{\bf r}_{\rm A}})\equiv\Delta E^{(1)}({{\bf r}_{\rm A}})
     = -\frac{\mu_0}{\pi}
     \sum_n \int_0^{\infty} \frac{\dif\omega}{\omega_n+\omega}
     \,{\bf d}_{0n}
\nonumber\\[1ex]&&\quad\times
     \Bigl\{\omega^2
     \Im ^\parallel\bm{G}^{(1)\parallel}({{\bf r}_{\rm A}},
     {{\bf r}_{\rm A}},\omega)
     -\,\omega_n\omega
\nonumber\\[1ex]&&\quad
     \times\bigl[
     \Im ^\perp\bm{G}^{(1)\perp}({{\bf r}_{\rm A}},
     {{\bf r}_{\rm A}},\omega)
     +\Im ^\perp\bm{G}^{(1)\parallel}({{\bf r}_{\rm A}},
     {{\bf r}_{\rm A}},\omega)
\nonumber\\[1ex]&&\quad
     +\,\Im ^\parallel\bm{G}^{(1)\perp}({{\bf r}_{\rm A}},
     {{\bf r}_{\rm A}},\omega) \bigr]
     \Bigr\} {\bf d}_{n0}.
\end{eqnarray}

To further evaluate this expression, it is convenient to express
it in terms of the whole scattering Green tensor rather than its
imaginary part. For this purpose, we write
\mbox{${\rm Im}\,\bm{G}^{(1)}$ $\!=$ $\!(\bm{G}^{(1)}$
$\!-$ $\!\bm{G}^{(1)\ast})/(2i)$}, recall the relation
(\ref{E20}), and change the integration variable
from $-\omega$ to $\omega$. Equation (\ref{E47}) then
changes to
\begin{widetext}
\begin{eqnarray}
\label{E48}
     \hspace{-3ex}
     U({{\bf r}_{\rm A}})
     &=& \frac{\mu_0}{2i\pi}
     \sum_n {\bf d}_{0n}
     \biggl(\int_0^{\infty}\frac{\dif \omega}{\omega_n+\omega}
     \Bigl\{ \omega_n\omega  \bigl[
     {}^\perp\bm{G}^{(1)\perp}({{\bf r}_{\rm A}},{{\bf r}_{\rm A}},\omega)
     + {}^\perp\bm{G}^{(1)\parallel}({{\bf r}_{\rm A}},
     {{\bf r}_{\rm A}},\omega)
     + {}^\parallel\bm{G}^{(1)\perp}({{\bf r}_{\rm A}},
     {{\bf r}_{\rm A}},\omega) \bigr]
\nonumber\\&&
     -\omega^2
     {}^\parallel\bm{G}^{(1)\parallel}({{\bf r}_{\rm A}},
     {{\bf r}_{\rm A}},\omega)
     \Bigr\}
     + \int^0_{-\infty}\frac{\dif \omega}{\omega_n-\omega}
     \Bigl\{ \omega_n\omega  \bigl[
     {}^\perp\bm{G}^{(1)\perp}({{\bf r}_{\rm A}},{{\bf r}_{\rm A}},\omega)
     + {}^\perp\bm{G}^{(1)\parallel}({{\bf r}_{\rm A}},
     {{\bf r}_{\rm A}},\omega)
     + {}^\parallel\bm{G}^{(1)\perp}({{\bf r}_{\rm A}},
     {{\bf r}_{\rm A}},\omega) \bigr]\nonumber\\
     &&+\omega^2
      {}^\parallel\bm{G}^{(1)\parallel}({{\bf r}_{\rm A}},
      {{\bf r}_{\rm A}},\omega)
     \Bigr\} \biggr) {\bf d}_{n0}\,.
\end{eqnarray}
\end{widetext}
This equation can be greatly simplified by using contour-integral
techniques. Note that 
$\bm{G}^{(1)}({{\bf r}_{\rm A}},{{\bf r}_{\rm A}},\omega)$
is an analytic function in the upper complex half plane
(\mbox{${\rm Im}\,\omega$ $\!\ge$ $\!0$},
\mbox{$\omega$ $\!\neq$ $\!0$}). Further,
$^\perp\bm{G}^{(1)\perp}({{\bf r}_{\rm A}},{{\bf r}_{\rm A}},\omega)$,
$^\perp\bm{G}^{(1)\|}({{\bf r}_{\rm A}},{{\bf r}_{\rm A}},\omega)$,
and $^\|\bm{G}^{(1)\perp}({{\bf r}_{\rm A}},{{\bf r}_{\rm A}},\omega)$
tend to zero as $\omega$ approaches zero, because the asymptote
of the Green tensor contains no transverse components
[cf. Eq.~(\ref{B4})]. Finally, the term
$\omega^2\,{^\|}\bm{G}^{(1)\|}({{\bf r}_{\rm A}},
{{\bf r}_{\rm A}},\omega)$ is also well-behaved for vanishing
$\omega$, as can be seen from Eq.~(\ref{B8}).
Consequently, the integrands of the $\omega$-integrals
in Eq.~(\ref{E48}) are analytic functions without
poles in the whole upper complex half plane, including the real axis.
We may therefore apply Cauchy's theorem, and
replace the integral over the positive (negative) real half axis
by a contour integral along the positive imaginary half
axis (introducing the purely imaginary coordinate
\mbox{$\omega$ $\!=$ $\!iu$}) and along a quarter circle with infinite
radius in the first (second) quadrant of the complex frequency
plane. Since the integrals along the infinitely large
quarter circles vanish [cf. Eq.~(\ref{B3})], we finally arrive at
\begin{equation}
\label{E49}
     U({{\bf r}_{\rm A}})
     = \frac{\mu_0}{\pi}
     \sum_n
     \int_0^{\infty}\!\! \dif u\,
     \frac{\omega_n u^2}{\omega_n^2 + u^2}
     {\bf d}_{0n} \bm{G}^{(1)}({{\bf r}_{\rm A}},
     {{\bf r}_{\rm A}},iu){\bf d}_{n0}\,,
\end{equation}     
where the identity $\bm{G}^{(1)}$ $\!=$ $\!{}^{\perp}\bm{G}^{(1)\perp}$
$\!+$ $\!{}^{\perp}\bm{G}^{(1)\parallel}$ $\!+$
$\!{}^{\parallel}\bm{G}^{(1)\perp}$
$\!+$ $\!{}^{\parallel}\bm{G}^{(1)\parallel}$ has been taken into account. 

Introducing the (lowest-order) ground-state polarizability tensor
\begin{equation}
\label{E50}
     \bm{\alpha}(\omega)= \lim_{\eta\to 0+}
     \frac{2}{\hbar}\sum_n
     \frac{\omega_n}{\omega_n^2-\omega^2-i\eta\omega}
     \,{\bf d}_{0n}\otimes{\bf d}_{n0}
\end{equation}
of the  atomic system (see, e.g., \cite{Davydov}),
we may represent Eq.~(\ref{E49}) in the equivalent form of
\begin{equation}
\label{E51}
     U({{\bf r}_{\rm A}})
     = \frac{\hbar\mu_0}{2\pi}
     \int_0^{\infty} \dif u \,u^2 \alpha_{ij}(iu)
     \,G^{(1)}_{ij}({{\bf r}_{\rm A}},{{\bf r}_{\rm A}},iu).
\end{equation}
It is worth noting that Eq.~(\ref{E51}) directly follows from QED in
causal media, without the need of additional assumptions borrowed
from other fields. Equation (\ref{E51})
expresses the vdW potential of an arbitrary atomic
system (such as an atom or a molecule) in the
presence of an arbitrary configuration of dispersing and
absorbing macroscopic dielectric bodies in terms of the
polarizability tensor of the atomic system in lowest order
of perturbation theory and the scattering
Green tensor of the macroscopic Maxwell equations.

In particular for an atom, one can make use of the spherical
symmetry and reduce Eq.~(\ref{E51}) to
\begin{equation}
\label{E52}
     U({{\bf r}_{\rm A}})
     = \frac{\hbar\mu_0}{2\pi}
     \int_0^{\infty} \dif u \,u^2 \alpha(iu)
     \,G^{(1)}_{ii}({{\bf r}_{\rm A}},{{\bf r}_{\rm A}},iu),
\end{equation}
where
\begin{equation}
\label{E53}
     \alpha(\omega) =
     \lim_{\eta\to 0+}
     \frac{2}{3\hbar}\sum_n
     \frac{\omega_n}{\omega_n^2-\omega^2-i\eta\omega}
     \,|{\bf d}_{0n}|^2.
\end{equation}
This result
agrees with the results inferred from \mbox{(se}\-m\mbox{i-)c}las\-sical
linear response theory. Note that the
field susceptibility introduced in Refs.~\cite{McLachlan}
and \cite{WylieSipe} differs from the scattering
Green tensor by a factor of $\omega^2$. Needless to say, that in
the special case of a two-level atom, Eq.~(\ref{E52})
reduces to the result, e.g., in Ref.~\cite{Argawal2}.
The derivation of Eq.~(\ref{E52}) shows that it can be
regarded as the natural extension of the QED results obtained
on the basis of the normal-mode formalism, which ignores
material absorption.

\subsection{The vdW energy in the multipolar-coupling scheme}
\label{Subsec:vdW_mulcf}

Let us turn to the multipolar-coupling scheme widely used
for studying the interaction of electromagnetic fields
with atoms and molecules. Just as in standard QED, so in
the present formalism the multipolar-coupling Hamiltonian can
be obtained from the minimal-coupling Hamiltonian by means
of a Power--Zienau transformation,
\begin{equation}
\label{E54}
\hat{\cal{H}} = \hat{U}^\dagger \hat{H} \hat{U},
\end{equation}
where
\begin{eqnarray}
\label{E55}
      \hat{U} &=&
      \exp\!\left[
      \frac{i}{\hbar}\int\! {\rm d}^3{r}\,
      \hat{{\bf P}}_{\rm A}({\bf r})\hat{\bf A}({\bf r})\right]
\end{eqnarray}
with
\begin{equation}
\label{E56}
      \hat{\bf P}_{\rm A}({\bf r}) = \sum_{\alpha}
      q_{\alpha}\hat{\bf r}_{\alpha}
      \int_0^1 {\rm d}\lambda \,
      \delta\!\left[{\bf r}\!-\!\left({{\bf r}_{\rm A}}\!+\!\lambda
      \hat{\bf r}_{\alpha} \right)\right]
\end{equation}
being the polarization associated with the (neutral)
atomic system. Using $\hat{H}$ from Eq.~(\ref{E3}), we derive
\cite{Knoll01,Ho02}
\begin{eqnarray}
\label{E57}
\lefteqn{
      \hat{\cal H} = \int\! {\rm d}^3{r} \int_0^\infty\! {\rm d}\omega
      \,\hbar\omega\,\hat{\bf f}^{\dagger}({\bf r},\omega)
      {}\hat{\bf f}({\bf r},\omega)
      + \sum_{\alpha}\frac{1}{2m_{\alpha}}
      \bigg\{\hat{{\bf p}}_{\alpha}
}
\nonumber\\[1ex]&& 
      +\, q_{\alpha}\int_0^1\!{\rm d}\lambda\,\lambda
      \hat{\bf r}_{\alpha}
      \times
      \hat{\bf B}\left[ {{\bf r}_{\rm A}}\!+\!\lambda
      \hat{\bf r}_{\alpha}
      \right]
      \bigg\}^2
\nonumber\\[1ex]&& 
      +\,\frac{1}{2\varepsilon_0}
      \!\int\! {\rm d}^3{r} \,
      \hat{\bf P}_{\rm A}({\bf r})  \hat{\bf P}_{\rm A}({\bf r})
      - \int\! {\rm d}^3{r}\,
      \hat{{\bf P}}_{\rm A}({\bf r})
      \hat{\bf E}_{\rm M}({\bf r}),
      \quad
      \end{eqnarray}
where \mbox{$\hat{\bf B}({\bf r})$ $\!=$
$\!\bm{\nabla}\!\times\!\hat{\bf A}({\bf r})$} with
$\hat{\bf A}({\bf r})$ from Eq.~(\ref{E8}) [together
with Eqs.~(\ref{E10}) and (\ref{E11})],
and $\hat{\bf E}_{\rm M}({\bf r})$ is defined by
Eq.~(\ref{E14}) [together with Eq.~(\ref{E11})].
Note that in the multipolar-coupling scheme the operator of the
electric field strength is defined according to
\begin{eqnarray}
\label{E58}
         \hat{\bf E}({\bf r})
         &=& - \frac{1}{i\hbar}
         \left[\hat{\bf A}({\bf r}),\hat{\cal H}\right]
         - \bm{\nabla} \hat{\varphi}_{\rm M}({\bf r})
         - \bm{\nabla} \hat{\varphi}_{\rm A}({\bf r})
\nonumber\\[1ex]
         &=& \hat{\bf E}_{\rm M}({\bf r})
         - \frac{1}{\varepsilon_0}\,\hat{\bf P}_{\rm A}({\bf r}),
\end{eqnarray}
i.e.,
\begin{eqnarray}
\label{E59}
       \varepsilon_0
       \hat{\bf E}_{\rm M}({\bf r})
       = \varepsilon_0
       \,\,\hat{\bf E}({\bf r})
       + \hat{{\bf P}}_{\rm A}({\bf r}).
\end{eqnarray}
Hence, $\varepsilon_0\hat{\bf E}_{\rm M}({\bf r})$ has the
meaning of the displacement field with respect to the
polarization of the atomic system.

In the electric-dipole approximation, Eq.~(\ref{E57})
simplifies to
\begin{equation}
\label{E60}
      \hat{\cal H} =  \hat{\cal H}_{\rm F}
      + \hat{\cal H}_{\rm A} + \hat{\cal H}_{\rm AF}\,,
\end{equation}
where
\begin{equation}
\label{E61}
      \hat{\cal H}_{\rm F}
      = \int {\rm d}^3{r} \int_0^\infty {\rm d} \omega
      \,\hbar\omega\,\hat{\bf f}^{\dagger}({\bf r},\omega)
      \hat{\bf f}({\bf r},\omega)
      \end{equation}
and
\begin{eqnarray}
\label{E62}
      \hat{\cal H}_{\rm A} = \sum_{\alpha}
      \frac{\hat{{\bf p}}_{\alpha}^2}{2m_{\alpha}}
      + \frac{1}{2\varepsilon_0}
      \int {\rm d}^3{r} \,
      \hat{\bf P}_{\rm A}({\bf r}) \hat{\bf P}_{\rm A}({\bf r}),
      \end{eqnarray}
respectively, are the unperturbed Hamiltonians of the
medium-assisted electromagnetic field and the atomic system, and
\begin{equation}
\label{E63}
      \hat{\cal H}_{\rm AF}
      = - \hat{\bf d}\hat{\bf E}_{\rm M}({{\bf r}_{\rm A}})
      \end{equation}
is the interaction energy between them, where $\hat{\bf d}$ is
the atomic dipole operator given by Eq.~(\ref{E31}). Recall
that $\hat{\bf E}_{\rm M}({{\bf r}_{\rm A}})$ must be thought
of a being expressed in terms of the fundamental field
variables $\hat{\bf f}({{\bf r}_{\rm A}})$ [and
$\hat{\bf f}^\dagger({{\bf r}_{\rm A}})$].
Comparing Eq.~(\ref{E62}) with Eq.~(\ref{E25}), we see that,
on taking into account the the relationship
\begin{equation}
\label{E63-1}
{\textstyle\frac{1}{2}}\int {\rm d}^3 r\,
\hat{\rho}_{\rm A}({\bf r})\hat{\varphi}_{\rm A}({\bf r})
= \frac{1}{2\varepsilon_0}\int {\rm d}^3 r\,
\hat{\bf P}^\parallel_{\rm A}({\bf r}) \hat{\bf P}^\parallel_{\rm A}({\bf r}),
\end{equation}
the atomic Hamiltonians $\hat{\cal H}_{\rm A}$ and
$\hat{H}_{\rm A}$ are different from each other, so that the
solution of the eigenvalue problem
\begin{equation}
\label{E27a}
       \hat{\cal H}_{\rm A}\rdirac{n'}=E_n'\rdirac{n'}
\end{equation}
may be different from that defined by Eq.~(\ref{E27}).
Keeping in mind this difference, we drop the primes denoting the
atomic eigenvalues and eigenstates from here on.

In contrast to the interaction energy in the
minimal-coupling scheme, Eq.~(\ref{E28}), the interaction energy
in the multipolar-coupling scheme, Eq.~(\ref{E63}),
is linear in $\hat{\bf f}({{\bf r}_{\rm A}})$ and
$\hat{\bf f}^\dagger({{\bf r}_{\rm A}})$. As a consequence of the
latter, there is no first-order correction to the ground
state energy. We thus have
\begin{equation}
\label{E64}
\Delta E \simeq \Delta_2 E,
\end{equation}
where
\begin{eqnarray}
\label{E65}
\Delta_2 E &=& \sum_n\int_0^{\infty}\dif\omega\int\dif^3{r}
\nonumber\\&&\times
      \;\frac{|\ldirac{0}\ldirac{\{0\}}
              \hat{\bf d}\hat{\bf E}_{\rm M}({\bf r})
             \rdirac{\{{\bf 1}({\bf r},\omega)\}}\rdirac{n}|^2}
      {E_0-(E_n+\hbar\omega)}
      \quad
\end{eqnarray}
[cf. the first line in Eq.~(\ref{E34}) with $\hat{\cal H}_{AF}$
instead of $\hat{H}_{\rm AF}^{\rm (I)}$]. In complete
analogy to the derivation of Eq.~(\ref{E41}), we find that
\begin{equation}
\label{E67}
     \Delta_2 E
     = - \frac{\mu_0}{\pi}\sum_n
     \int_0^{\infty}\!\!\dif\omega\,
     \frac{\omega^2}{\omega_n+\omega}
     \,{\bf d}_{0n}
     {\rm Im}\,\bm{G} ({{\bf r}_{\rm A}},{{\bf r}_{\rm A}},\omega)
     {\bf d}_{n0}\,.    
\end{equation}

To further evaluate the ${{\bf r}_{\rm A}}$-dependent part
$U({\bf r}_{\rm A})$ $\!=$ $\!\Delta_2^{(1)}E({\bf r}_{\rm A})$
of $\Delta_2 E$, which results from the scattering part of the
Green tensor, $\bm{G}^{(1)}({{\bf r}_{\rm A}},{{\bf r}_{\rm A}},\omega)$,
and gives the vdW energy, we again
write \mbox{${\rm Im}\,\bm{G}^{(1)}$ $\!=$ $\!(\bm{G}^{(1)}$
$\!-$ $\!\bm{G}^{(1)\ast})/(2i)$}, use the relation (\ref{E20}),
and change the integration variable from $-\omega$ to $\omega$.
After some algebra we arrive at
\begin{eqnarray}
\label{E68}
     U({{\bf r}_{\rm A}})
     &=& -\frac{\mu_0}{2i\pi}
     \sum_n {\bf d}_{0n}
     \biggl[\int_0^{\infty}\dif \omega
     \frac{\omega^2}{\omega_n+\omega}
     \,\bm{G}^{(1)}({{\bf r}_{\rm A}},{{\bf r}_{\rm A}},\omega)
\nonumber\\&&
     - \int^0_{-\infty}\dif \omega
     \frac{\omega^2}{\omega_n-\omega}
     \,\bm{G}^{(1)}({{\bf r}_{\rm A}},{{\bf r}_{\rm A}},\omega)
     \biggr] {\bf d}_{n0}\,.
     \qquad
\end{eqnarray}
As we already know, the integrands of the two frequency integrals
appearing in Eq.~(\ref{E68}) are analytic functions in
the upper half of the complex frequency plane, including the
real axis [cf.
Eq.~(\ref{B8})].
We therefore
can apply contour integral techniques in a similar way as in
the derivation of Eq.~(\ref{E49}) from Eq.~(\ref{E48}).
It is not difficult to see that the result reads
\begin{equation}
\label{E69}
     U({{\bf r}_{\rm A}})
     = \frac{\mu_0}{\pi}
     \sum_n
     \int_0^{\infty}\!\! \dif u\,
     \frac{\omega_n u^2}{\omega_n^2 + u^2}
     {\bf d}_{0n} \bm{G}^{(1)}({{\bf r}_{\rm A}},
     {{\bf r}_{\rm A}},iu){\bf d}_{n0}\,,
\end{equation}
which has exactly the same form as the
minimal-cou\-p\-l\-ing result (\ref{E49}), so that it can
also be given in the form of Eq.~(\ref{E51}). Recall that the
values of $\omega_n$ and ${\bf d}_{0n}$ obtained in the
minimal-coupling scheme may be different from those obtained in the
multipolar-coupling scheme, because of the somewhat
different eigenvalue equations (\ref{E27}) and (\ref{E27a}).

\section{Application: an atom near a sphere}
\label{Sec:Appl}

Let us apply the theory to an atom near a dispersing and
absorbing dielectric (micro-)sphere surrounded by vacuum.
The material of the sphere of radius $R$ is assumed to be
homogeneous and isotropic, having a permittivity 
$\varepsilon(\omega)$. The coordinate system is chosen such
that its origin lies at the center of the sphere.
The scattering Green tensor can be given by \cite{Li94}
\begin{eqnarray}
\label{E70}
\lefteqn{
 \bm{G}^{(1)}({{\bf r}_{\rm A}},{{\bf r}_{\rm A}},iu)
 =\frac{u}{4\pi c}\sum_{n=1}^{\infty}
 \sum_{m=0}^{n}
 \!\left(2-\delta_{m0}\right)
 \frac{2n+1}{n(n+1)}
}
\nonumber\\[1ex]&&\times\;
 \frac{(n-m)!}{(n+m)!}
 \biggl[B^M_n\sum_{p=-1,1}{\bf M}_{nm,p}({{\bf r}_{\rm A}})
 \otimes{\bf M}_{nm,p}({{\bf r}_{\rm A}})
\nonumber\\[1ex]&&\hspace{4ex}
 +\,B^N_n\sum_{p=-1,1}{\bf N}_{nm,p}({{\bf r}_{\rm A}})
 \otimes{\bf N}_{nm,p}({{\bf r}_{\rm A}})\biggr],
\end{eqnarray}
where ${\bf M}_{nm,p}({{\bf r}_{\rm A}})$ and ${\bf N}_{nm,p}
({{\bf r}_{\rm A}})$ are even \mbox{($p$ $\!=$ $\!1$)} and
odd \mbox{($p$ $\!=$ $\!-1$)} spherical wave vector functions,
which can be expressed in terms of spherical Hankel
functions of the first kind, $h_n^{(1)}(r)$, and associated
Legendre functions, $P_n^m(\cos\theta)$, in the following way:
\begin{eqnarray}
\label{E71}
\lefteqn{
 {\bf M}_{nm,\pm1}({{\bf r}_{\rm A}})
 =\mp\frac{m}{\sin(\theta)}\,h_n^{(1)}(k_0 r_{\rm A}) P_n^m(\cos\theta)
}
\nonumber\\[1ex]&&\times\;
 \begin{array}{c}\sin\\
 \cos
 \end{array}
 (m\phi){\bf e}_{\theta}
 -h_n^{(1)}(k_0 r_{\rm A})\,
 \frac{\dif P_n^m(\cos\theta)}{\dif\theta}\,
 \begin{array}{c}\cos\\
 \sin
 \end{array}
 (m\phi){\bf e}_{\phi},
\nonumber\\&& 
\\
\label{E72}
\lefteqn{
 {\bf N}_{nm,\pm1}({{\bf r}_{\rm A}})
 =n(n+1)\,\frac{h_n^{(1)}(k_0 r_{\rm A})}{k_0 r_{\rm A}}\,P_n^m(\cos\theta)
}
\nonumber\\&&\times\;
 \begin{array}{c}
 \cos\\
 \sin
 \end{array}(m\phi){\bf e}_r
 +\frac{1}{k_0 r_{\rm A}}
 \frac{\dif[r_{\rm A}h_n^{(1)}(k_0 r_{\rm A})]}{\dif r_{\rm A}}
 \biggl[\frac{\dif P_n^m(\cos\theta)}{\dif\theta}
\nonumber\\&&\times\;
 \begin{array}{c}
 \cos\\
 \sin
 \end{array}
 (m\phi){\bf e}_{\theta}
 \mp\frac{m}{\sin\theta}\,P_n^m(\cos\theta)
 \begin{array}{c}
 \sin\\
 \cos
 \end{array}
 (m\phi){\bf e}_{\phi}\biggr].
\end{eqnarray}
Here, \mbox{$k_0$ $\!=$ $\!\omega/c$ $\!=$ $\!iu/c$} is the
vacuum wave number, and ${\bf e}_r$, ${\bf e}_{\theta}$,
${\bf e}_{\phi}$, are the mutually orthogonal unit vectors
pointing in the directions of $r$, $\theta$, and $\phi$,
respectively. The coefficients $B^M_n$ and $B^N_n$ in
Eq.~(\ref{E70}) read
\begin{eqnarray}
\label{E73}
\lefteqn{
      B^M_n = B^M_n(iu)
}
\nonumber\\[1ex]&&
       = - \frac
       {\bigl[ z_1j_n(z_1)\bigr]' j_n(z_0)
       - \bigl[ z_0j_n(z_0)\bigr]' j_n(z_1) }
       {\bigl[ z_1j_n(z_1)\bigr]' h_n^{(1)}(z_0)
       -  \bigl[z_0 h_n^{(1)}(z_0)\bigr]' j_n(z_1) }\,,
       \quad
\end{eqnarray}
\begin{eqnarray}
\label{E74}
\lefteqn{
      B^N_n = B^N_n(iu)
}
\nonumber\\[1ex]&&
       = -  \frac
       { \varepsilon(iu)
       j_n(z_1) \bigl[z_0 j_n(z_0)\bigr]'
       - j_n(z_0) \bigl[z_1 j_n(z_1)\bigr]' }
       { \varepsilon(iu)
       j_n(z_1) \bigl[ z_0 h_n^{(1)}(z_0)\bigr]'
       -h_n^{(1)}(z_0)\bigl[z_1 j_n(z_1)\bigr]' }\,,
       \quad
\nonumber\\&&
\end{eqnarray}
where \mbox{$z_0$ $\!=$ $\!k_0 R$} and \mbox{$z_1$ $\!=$ $\!k R$},
with $k$ $\!=$ $\!k_0\sqrt{\varepsilon(iu)}$ being
the wave number inside the sphere, and $j_n(z)$ is the
spherical Bessel function of the first kind. The primes indicate
differentiations with respect to $z_0$ or $z_1$, respectively.
The coefficients $B^M_n$ represent contributions from
transverse electric (TE) waves reflected at the surface of the
sphere, while the coefficients $B^N_n$ represent those
from transverse magnetic (TM) waves.

Substituting the trace of $\bm{G}^{(1)}({\bf r}_{\rm A},
{\bf r}_{\rm A},\omega)$ from Eq.~(\ref{E70}) [together with
Eqs.~(\ref{E71}) and (\ref{E72})] into Eq.~(\ref{E52}) yields
the vdW energy sought. The sums over $p$ can
then easily be performed using the orthogonality of the unit
vectors ${\bf e}_r$, ${\bf e}_{\theta}$, and ${\bf
e}_{\phi}$, and the sum over $m$ can be performed with the
aid of the summation formulas
in Appendix \ref{AppC}. So after a lengthy, but straightforward
calculation we arrive at the following result:
\begin{eqnarray}
\label{E75}
\lefteqn{
    U({\bf r}_{\rm A})
    = - \frac{\hbar\mu_0}{8\pi^2 c}
    \int_0^{\infty}\dif u
    \Biggl(
    u^3 \alpha(iu)\sum_{n=1}^{\infty}(2n+1)
}
\nonumber\\&&\times\,
    \Biggl\{ B^M_n \left[h^{(1)}_n(k_0 r_{\rm A})\right]^2
    + n(n+1)B^N_n \left[\frac{h^{(1)}_n(k_0 r_{\rm A})}
    {k_0 r_{\rm A}}\right]^2
\nonumber\\&&\hspace{4ex}
    +\,B^N_n \left[\frac{1}{k_0 r_{\rm A}}
    \frac{\dif[r_{\rm A} h^{(1)}_n(k_0 r_{\rm A})]}
    {\dif r_{\rm A}}\right]^2
    \Biggr\}\Biggr).
\end{eqnarray}
Note that the vdW potential does not depend on the
angle variables of the atomic position, but only on the
distance of the atom from the center of the sphere, as can be
anticipated from the symmetry of the system. Recall that
the terms proportional to $B^{M(N)}_n$ represent the contributions
from the TE (TM) waves. Equation (\ref{E75}) applies to an
arbitrary dielectric sphere. In particular, when material
absorption is omitted, then the result in Ref.~\cite{MarvinToigo}
can be recovered.

\subsubsection{Long-distance limit}
\label{Long_range}

A detailed analysis of Eq.~(\ref{E75}) requires numerical
computation. Here, however, we would like to focus our attention on two
interesting limiting cases, where the atom is very far from or
very close to the sphere. Let us first consider the limit of the atom being 
far away from the sphere,
\begin{equation}
\label{E76}
    r_{\rm A}\gg R.
\end{equation}
In this case, Eq.~(\ref{E75}) reduces to
\begin{eqnarray}
\label{E77}
\lefteqn{
    U({\bf r}_{\rm A})
    \simeq -\frac{\hbar cR^3}{4\pi^2\varepsilon_0}
    \frac{1}{r_{\rm A}^7}
    \int_0^\infty \dif z\,
    \alpha(icz/r_{\rm A})\,
    \frac{\varepsilon(icz/r_{\rm A})-1}{\varepsilon(icz/r_{\rm A})+2}
}
\nonumber\\[1ex]&&\hspace{10ex}\times\,
    \left[2\left(1+z\right)^2+\left(1+z+z^2\right)^2\right]e^{-2z}
    \qquad
\end{eqnarray}
%
(see Appendix \ref{AppD}), where it turns out that the TE waves
do not contribute. Since the inequalities 
\mbox{$\varepsilon(icz/r_{\rm A})$ $\!>$ $\!1$}
and \mbox{$\alpha(icz/r_{\rm A})$ $\!>$ $\!0$} are valid,
the vdW potential is negative, and the
resulting force between the atom and the sphere is attractive. 

As is seen from Eq.~(\ref{E77}), the main contribution to the integral
comes from the region where \mbox{$z\lesssim 1/2$}. Therefore,
for sufficiently large distances, the contributions from small
frequencies dominate, and we can (approximately) replace
the atomic polarizability and the material permittivity
in Eq.~(\ref{E77}) with their static values
\mbox{$\alpha^{(0)}$ $\!=$ $\!\alpha(\omega$ $\!=$ $\!0)$}
and \mbox{$\varepsilon^{(0)}$ $\!=$ $\!\varepsilon(\omega$ $\!=$ $\!0)$},
respectively. The integration can then be
performed in closed form to yield the asymptotic distance law
\begin{eqnarray}
\label{E78}
    U({\bf r}_{\rm A})
    = -\frac{23\hbar cR^3\alpha^{(0)}}{16\pi^2\varepsilon_0}
    \frac{\varepsilon^{(0)}-1}{\varepsilon^{(0)}+2}\,
    \frac{1}{r_{\rm A}^7}
    \quad
    \left(\frac{r_A}{R}\to\infty\right),
    \quad
\end{eqnarray}
in this so-called retarded limit. Note that in the opposite nonretarded limit,
where the contributions of $\alpha(\omega)$ and $\varepsilon(\omega)$
at all frequencies have to be retained, Eq.~(\ref{E77}) reduces to the 
result given in Ref.~\cite{MarvinToigo}, where a $r_{\rm A}^{-6}$ law
was found. The (formal) limit \mbox{$\varepsilon^{(0)}$
$\!\to$ $\!\infty$} in Eq.~(\ref{E78}) obviously corresponds to a metallic sphere
\begin{eqnarray}
\label{E78.1}
    U({\bf r}_{\rm A})
    = -\frac{23\hbar cR^3\alpha^{(0)}}{16\pi^2\varepsilon_0}
    \frac{1}{r_{\rm A}^7}
    \quad
    \left(\frac{r_A}{R}\to\infty\right).
    \quad
\end{eqnarray}
Note that the decrease of the force with the distance is
of three powers stronger than in the case of the atom
being near a planar body.

In particular, if we introduce the
static polarizability of the sphere (see, e.g., \cite{Jackson})
\begin{equation}
\label{E79}
    \alpha_{\mathrm{sph}}^{(0)}=
    4\pi\varepsilon_0
    \,\frac{\varepsilon^{(0)}-1}{\varepsilon^{(0)}+2}\,R^3,
\end{equation}
we may rewrite Eq.~(\ref{E78}) as
\begin{eqnarray}
\label{E80}
    U({\bf r}_{\rm A})
    = -\frac{\alpha^{(0)}_{\mathrm{sph}}\alpha^{(0)}}
    {(4\pi\varepsilon_0)^2}
    \,\frac{23\hbar c}{4\pi}\, \frac{1}{r_{\rm A}^7}
    \quad
    \left(\frac{r_A}{R}\to\infty\right).
    \quad
\end{eqnarray}
Interestingly, Eq.~(\ref{E80}) also applies to the
vdW potential between two atoms \cite{CasimirPolder},
if the (static) polarizability of the sphere is replaced with
the polarizability of the second atom.

\subsubsection{Short-distance limit}
\label{Short_range}

\noindent
Let us now proceed to the short-distance limit of the atom
being located at a position very close to the sphere, i.e.,
\begin{equation}
\label{E81}
     \frac{\Delta r_{\rm A}}{R} \ll 1
\end{equation}
($\Delta r_{\rm A}$ $\!\equiv$ $\!r_{\rm A}$ $\!-$ $\!R$).
In this case, from Eq.~(\ref{E75}) it follows that
\begin{equation}
\label{E82}
   U({\bf r}_{\rm A})
   \simeq
   -\frac{\hbar}{16\pi^2\varepsilon_0}
   \,\frac{1}{(\Delta r_{\rm A})^3}
   \int_0^{\infty}\!\dif u\,\alpha(iu)
   \,\frac{\varepsilon(iu)-1}{\varepsilon(iu)+1}
\end{equation}
(see Appendix \ref{AppE}). Note that again
the TE waves do not contribute to $U({\bf r}_{\rm A})$.

As expected, the dependence on distance of the (attractive)
vdW potential corresponds to that obtained in the
case of the atom being near a planar body.
In fact, it exactly looks like that derived
in Ref.~\cite{Zhou} for an atom in the vicinity of a
planar, semi-infinite, non-absorbing dielectric.
In particular, if the (model) assumption
$[\varepsilon(iu)$ $\!-$ $\!1]/[\varepsilon(iu)$ $\!+$ $\!1]$
$\!=$ $\!1$ were made for all values of $u$, then
Eq.~(\ref{E82}) would lead, on using Eq.~(\ref{E53}),
to the result \cite{CasimirPolder}
\begin{eqnarray}
\label{E83}
   U({\bf r}_{\rm A})
   \simeq -\frac{\langle 0|\hat{\bf d}^2|0\rangle}
   {48\pi\varepsilon_0}\,
   \frac{1}{(\Delta r_{\rm A})^3}
   \,.
\end{eqnarray}

\subsubsection{Material absorption}
\label{absorption}

To explore the effect of material absorption, we
may assume a permittivity of Drude-Lorentz type,
\begin{equation}
\label{E84}
    \varepsilon(\omega)=1+\sum_l
    \frac{\Omega_{l}^2}{\omega_{l}^2-\omega^2-i\omega\gamma_l}\,,
\end{equation}
where $\omega_{l}$ and $\gamma_l$
are respectively the (transverse) resonance frequencies
and the associated absorption constants, and
the frequencies $\Omega_{l}$ are proportional to the
so-called oscillator strengths.
From Eq.~(\ref{E84}) it is seen that in the limit
\mbox{$\omega$ $\!\to$ $\!0$} the resulting static permittivity
\begin{equation}
\label{E85}
    \varepsilon^{(0)}= 1+\sum_l
    \frac{\Omega_{l}^2}{\omega_{l}^2}
\end{equation}
is independent of the absorption parameters.
Since it is the static permittivity that enters Eq.~(\ref{E78}),
we see that the long-distance asymptote of the vdW
potential is not influenced by material absorption.

With decreasing distance the range of frequency that must be
taken into account increases. Thus, the frequency response
of the permittivity becomes crucial to the strength of
the vdW force. Let us consider the short-distance
law (\ref{E82}). From Eq.~(\ref{E84}) it follows that
\begin{equation}
\label{E86}
\frac{\partial}{\partial \gamma_l}
\frac{\varepsilon(iu)-1}{\varepsilon(iu)+1}
= - \frac{1}{[\varepsilon(iu)+1]^2}
\frac{2u\Omega_{l}^2}{(\omega_{l}^2+u^2+u\gamma_l)^2}\,,
\end{equation}
that is to say,
\begin{equation}
\label{E87}
\frac{\partial}{\partial \gamma_l}
\frac{\varepsilon(iu)-1}{\varepsilon(iu)+1}
< 0
\quad\mbox{if}\quad
u > 0.
\end{equation}
With regard to Eq.~(\ref{E82}), we therefore find that,
on recalling that \mbox{$\alpha(iu)$ $\!>$ $\!0$},
%
\begin{equation}
\label{E88}
\frac{\partial}{\partial \gamma_l}
\left|\frac{\partial U({\bf r}_{\rm A})}{\partial r_{\rm A}}\right|
< 0.
\end{equation}
Hence, the vdW force monotonically decreases with
increasing absorption constants.

\begin{figure}[htb]
\noindent
\includegraphics[width=1.\linewidth]{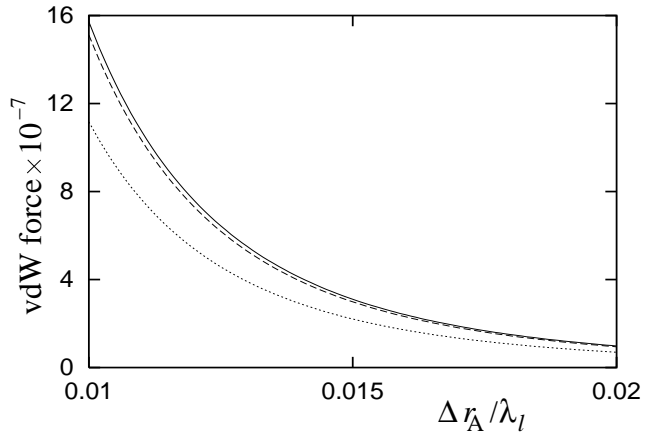}
\caption{
Absolute value of the normalized van der Waals force 
$C|\partial U({\bf r}_{\rm A})/\partial r_{\rm A} |$
[$C$ $\!=$ $\!16\pi^2\varepsilon_0/(\sum_n|{\bf d}_{0n}|^2\lambda_l^4)$,
$\lambda_l$ $\!=$ $\!2\pi c/\Omega_l$]
as a function of the atom-surface distance for
various strengths of material absorption. In the calculation, a metallic
permittivity of Drude type according to Eq.~(\ref{E84}) and a 
(degenerate) single-resonance atomic polarizability are assumed 
with
\mbox{$\omega_l$ $\!=$ $\!0$},
$\omega_n/\Omega_l$ $\!=$ $\!7\times10^{-1}$, and
$\gamma_l/\Omega_l$ $\!=$ $\!10^{-2}$ (solid line),
\mbox{$\gamma_l/\Omega_l$ $\!=$ $\!10^{-1}$} (dashed line),
and $\gamma_l/\Omega_l$ $\!=$ $\!1$ (dotted line). 
}
\label{force}
\end{figure}

{F}igure \ref{force} illustrates the
influence of material absorption on the vdW force acting on
an atom located near a metallic sphere
in the short-distance limit. It can be seen that the
effect of material absorption increases with
decreasing atom-surface distance. In particular, at a distance of
$\Delta r_{\rm A}$ $\!\simeq$
$10^{-2}\lambda_l$,
an increase of the relative absorption parameter from
$\gamma_l/\Omega_l$ $\!=$ $\!10^{-2}$
to
$\gamma_l/\Omega_l$ $\!=$ $\!1$
would reduce the magnitude of the force by nearly thirty percents.

\section{Conclusions}
\label{Sec:Concl}

Within the frame of macroscopic QED, we have derived an
expression for the vdW potential of an atomic system near
an arbitrary configuration of dispersing and absorbing bodies.
It generalizes the results obtained by means of normal-mode
expansion and may be regarded as a foundation of the
results inferred from linear response theory.
We have performed the calculations for both the
minimal-coupling scheme and the multipolar-coupling
scheme and shown that the results are essentially the
same.

We have applied the theory
to the vdW interaction between an atom and a sphere.
From the integral expression, we have derived the correct
long-distance law corresponding to the retardation limit
and recovered the short-distance law corresponding
to the non-retardation limit.
In particular, replacing in the long-distance law
the polarizability of the sphere with that of an atom
just yields the vdW potential between two atoms.
On the other hand, for sufficiently small distances of the
atom from the sphere the vdW potential approaches
the potential observed for an atom near a planar body.

It is worth noting that
in the long-distance limit it is the static permittivity
that enters the vdW potential. Hence material
absorption has no effect on it. However, with decreasing
distance of the atom from the sphere the relevant frequencies
extend for a finite (increasing) interval and material
absorption becomes substantial, thereby diminishing the
strength of the force.

In this article, we have restricted our attention to
ground-state systems and calculated the vdW potential
in lowest-order of perturbation theory with respect
to the interaction of the atomic system with the
medium-assisted electromagnetic field.
The theory allows of course extensions in several respects.
As a consequence of the lowest-order perturbation
theory, the energy denominators that enter the polarizability of
the atomic system are the unperturbed ones, without consideration
of the level shift and broadening caused by the presence
of the bodies. In fact, the polarizability of an atomic system
is expected to drastically change when it becomes close to
a macroscopic body and the spontaneous decay thus becomes
purely radiationless, with the decay rate being proportional
to $\Delta r_{\rm A}^{-3}$ \cite{Ho01}. Since the
level broadening is essentially determined by the
spontaneous-decay rate, the polarizability becomes
distance-dependent -- an effect that needs careful
consideration.

Since the electromagnetic field in (linear) magnetic
media can be quantized analogously \cite{Knoll01},
another interesting extension of the theory
be the inclusion in it of composite materials
characterized by both a complex permittivity and a
complex permeability. Interestingly, such materials,
which have been fabricated recently, are left-handed.
Last not least the underlying quantization scheme
renders it also possible to extend the theory
to atoms and molecules in excited states and
treat the motion of driven atomic systems.

\begin{acknowledgments}
We would like to thank Ludwig Kn\"{o}ll for valuable discussions.
This work was supported by the Deutsche Forschungsgemeinschaft.
\end{acknowledgments}


\appendix
\section{Derivation of Eq.~(\ref{E39})}
\label{Sec:sumrule}

From the commutation relation
\begin{equation}
\label{A1}
    \left[\hat{r}_{\alpha,i},\hat{H}_A\right]
    =\biggl[\hat{r}_{\alpha,i}\,,
    \sum_{\beta} \frac{\hat{{\bf p}}_{\beta}^2}{2m_{\beta}}\biggr]
    =\frac{i\hbar}{m_{\alpha}}\,\hat{p}_{\alpha,i}
\end{equation}
together with the eigenvalue equation (\ref{E27})
we find that
\begin{eqnarray}
\label{A2}
     \ldirac{0}\hat{p}_{\alpha,i}\rdirac{n}
     &\!=&\!
     -\frac{im_{\alpha}}{\hbar}\ldirac{0}
     \big[\hat{r}_{\alpha,i},\hat{H}_A\big]\rdirac{n}
\nonumber\\[1ex]
     &\!=&\!
     -\frac{im_{\alpha}}{\hbar}\ldirac{0}\hat{r}_{\alpha,i}\hat{H}_A
     -\hat{H}_A\hat{r}_{\alpha,i}\rdirac{n}
\nonumber\\[1ex]
     &\!=&\!
     -\frac{im_{\alpha}}{\hbar}(E_n-E_0)
     \ldirac{0}\hat{r}_{\alpha,i}\rdirac{n}
\nonumber\\[1ex]
     &\!=&\!
     -im_{\alpha}\omega_n\ldirac{0}\hat{r}_{\alpha,i}\rdirac{n}.
\end{eqnarray}
Thus
\begin{eqnarray}
\label{A3}
      \sum_{\alpha}\frac{q_{\alpha}}{m_{\alpha}}
      \ldirac{0}\hat{p}_{\alpha,i}\rdirac{n}
      &\!=&\!
      -i\omega_n\sum_{\alpha}q_{\alpha}\ldirac{0}
      \hat{r}_{\alpha,i}\rdirac{n}
\nonumber\\[1ex]
      &\!=&\!
      -i\omega_n\ldirac{0}\hat{d}_i\rdirac{n}.
\end{eqnarray}
Using Eq.~(\ref{A3}), we derive
\begin{eqnarray}
\label{A4}
\lefteqn{
 \frac{1}{2\hbar}\sum_n\omega_n
  \left(\ldirac{0}\hat{d}_i\rdirac{n}\ldirac{n}\hat{d}_j\rdirac{0}
  +\ldirac{0}\hat{d}_j\rdirac{n}\ldirac{n}\hat{d}_i\rdirac{0}\right)
}
\nonumber \\[1ex]&&\hspace{2ex}
 =\frac{i}{2\hbar}\sum_{\alpha}\frac{q_{\alpha}}{m_{\alpha}}
  \sum_n\left(\ldirac{0}\hat{p}_{\alpha,i}\rdirac{n}\ldirac{n}
  \hat{d}_j\rdirac{0}\right.
\nonumber  \\&&\hspace{20ex}
  \left.- \ldirac{0}\hat{d}_j\rdirac{n}\ldirac{n}
  \hat{p}_{\alpha,i}\rdirac{0}\right)
  \qquad
\nonumber\\[1ex]&&\hspace{2ex}
 =\frac{i}{2\hbar}\sum_{\alpha}\frac{q_{\alpha}}{m_{\alpha}}
  \ldirac{0}\big[\hat{p}_{\alpha,i},\hat{d}_j\big]\rdirac{0}
\nonumber\\[1ex]&&\hspace{2ex}
 =\frac{i}{2\hbar}\sum_{\alpha}\frac{q_{\alpha}}{m_{\alpha}}
  \ldirac{0}\Big[\hat{p}_{\alpha,i},
  \sum_{\beta}q_{\beta}\hat{r}_{\beta,j}\Big]\rdirac{0}
\nonumber\\[1ex]&&\hspace{2ex}
 =\sum_{\alpha}\frac{q_{\alpha}^2}{2m_{\alpha}}\delta_{ij},
\end{eqnarray}
which is just Eq.~(\ref{E39}).

\section{Asymptotic behavior of the Green tensor}
\label{AppB}

The asymptotic behavior of the Green tensor for large
frequencies reads \cite{Knoll01}
\begin{equation}
\label{B1}
     \lim_{|\omega|\rightarrow \infty}
     \frac{\omega^2}{c^2} \bm{G}({\bf r},{\bf r}',\omega)
     =-\bm{\delta}({\bf r}-{\bf r}'),
\end{equation}
\begin{equation}
\label{B2}
     \lim_{|\omega|\rightarrow \infty}
     \frac{\omega^2}{c^2} \bm{G}^{(0)}({\bf r},{\bf r}',\omega)
     =-\bm{\delta}({\bf r}-{\bf r}').
\end{equation}
If ${\bf r}$ and ${\bf r}'$ lie in a common region of constant permittivity,
we can use Eq. (\ref{E45}), and subtract
the two equations (\ref{B1}) and (\ref{B2}) to obtain
\begin{equation}
\label{B3}
     \lim_{|\omega|\rightarrow \infty}
     \frac{\omega^2}{c^2} \bm{G}^{(1)}({\bf r},{\bf r}',\omega)
     =0.
\end{equation}

In the low-frequency limit we have \cite{Knoll01}
\begin{equation}
\label{B4}
     \lim_{|\omega|\rightarrow 0}\frac{\omega^2}{c^2}
     \bm{G}({\bf r},{\bf r}',\omega)
     = - {^\|}\bm{L}^{-1}{^\|} ({\bf r},{\bf r}'),
\end{equation}
where
\begin{equation}
\label{B4-1}
     \bm{L}({\bf r},{\bf r}')
     = \lim_{|\omega|\rightarrow 0}
     \int\dif^3{s}\, \bm{\delta}^\parallel({\bf r}-\bm{s})
     \varepsilon({\bf s},\omega)
     \bm{\delta}^\parallel(\bm{s}-{\bf r}').
\end{equation}
Recalling that as \mbox{$|\omega|$ $\!\to$ $\!0$},
\begin{equation}
\label{B6}
\varepsilon({\bf r},\omega) \sim
\left\{
\begin{array}{l@{\quad}l}
\omega^0 & \mbox{for dielectrics},\\[1ex]
(i\omega)^{-1} & \mbox{for metals}
\end{array}
\right.
\end{equation}
[cf. Eq.~(\ref{E84})], from Eqs.~(\ref{B4}) and (\ref{B4-1})
we see that
\begin{equation}
\label{B8}
\lim_{|\omega|\to 0}\omega^2\bm{G}({\bf r},{\bf r}',\omega) =
M, \qquad M<\infty.
\end{equation}
Needless to say, that Eq.~(\ref{B8}) is also valid for the
scattering part of the Green tensor,
$\bm{G}^{(1)}({\bf r},{\bf r}',\omega)$.


\section{Summation formulas for Legendre polynomials}
\label{AppC}

The Legendre polynomials obey the relation \cite{Abramowitz}
\begin{equation}
\label{C1}
 \sum_{m=0}^n C_{nm}
 \cos(m\lambda)P_n^m(x)P_n^m(y)
 = P_n(\xi),
\end{equation}
where
\begin{eqnarray}
\label{C1.1}
&\displaystyle
  C_{nm} = \left(2\!-\!\delta_{m0}\right) \frac{(n\!-\!m)!}{(n\!+\!m)!},
\\
\label{C2}
&\displaystyle
 \xi\equiv xy+\sqrt{(1-x^2)(1-y^2)}\,\cos\lambda.
\end{eqnarray}
For
\mbox{$\lambda$ $\!=$ $\!0$} and
\mbox{$x$ $\!=$ $\!y$ $\!=$ $\!\cos\theta$},
Eq.~(\ref{C1}) reduces to
\begin{equation}
\label{C3}
 \sum_{m=0}^n
 C_{nm}P_n^m(\cos\theta)^2=1.
\end{equation}
Differentiating Eq.~(\ref{C1}) twice with respect
to $\lambda$ and putting \mbox{$\lambda$ $\!=$ $\!0$} and
\mbox{$x$ $\!=$ $\!y$ $\!=$ $\!\cos\theta$} afterwards yield
\begin{equation}
\label{C4}
 \sum_{m=0}^n
 C_{nm}\frac{m^2}{\sin^2\theta}\,P_n^m(\cos\theta)^2
 =\frac{n(n+1)}{2}\,.
\end{equation}
Finally, subsequent differentiations of Eq.~(\ref{C1})
with respect to $x$ and $y$
and again putting \mbox{$\lambda$ $\!=$ $\!0$} and
\mbox{$x$ $\!=$ $\!y$ $\!=$ $\!\cos\theta$} afterwards yield
\begin{equation}
\label{C5}
\sum_{m=0}^n C_{nm}
  \left[\frac{\dif P_n^m(\cos\theta)}{\dif\theta}\right]^2
  =\frac{n(n+1)}{2}\,.
\end{equation}

\section{Derivation of Eq.~(\ref{E77})}
\label{AppD}

In Eq.~(\ref{E75}), the spherical Hankel functions
$h^{(1)}_n(k_0r_{\rm A})$ $\!=$
$\!h^{(1)}_n(iur_{\rm A}/c)$ can be written in the form of \cite{Abramowitz}
\begin{equation}
\label{hexpansion}
       h^{(1)}_n(k_0r_{\rm A})
      =\sum_{j=0}^n h_j e^{-ur_{\rm A}/c}
      \left(\frac{c}{ur_{\rm A}}\right)^{j+1}
\end{equation}
with some complex coefficients $h_j$. From inspection of
Eqs.~(\ref{E73}) and (\ref{E74}) it is seen that $B^{M}_n$ and $B^{N}_n$
can be expanded in powers of $u$ at \mbox{$u$ $\!=$ $\!0$},
\begin{eqnarray}
\label{ABexpansion}
    B^{M,N}_n=\sum_{j=0}^{\infty}b^{M,N}_j
    \left(\frac{uR}{c}\right)^j.
\end{eqnarray}
From
Eqs.~(\ref{hexpansion}) and (\ref{ABexpansion})
it then follows that the integrand
of the (imaginary) frequency integral in Eq. (\ref{E75})
is a sum of terms, which are all of the same general structure
\begin{equation}
\label{CC3}
    f_{jk}(u)=\alpha(iu)u^3\left(\frac{uR}{c}\right)^j
    \left(\frac{c}{ur_{\rm A}}\right)^{k+2}e^{-2ur_{\rm A}/c}
\end{equation}
($j,k$ are nonnegative integers).
For $j$ $\!>$ $\!(k$ $\!-$ $\!1)$ this is a polynomial in
$u$ times an exponentially decaying function. The only relevant
contributions to the frequency integral come from the maximum
of $f_{jk}(u)$ at a frequency $u_0$ satisfying
\begin{equation}
\label{CC3-1}
\left.\frac{\dif}{\dif u}f_{jk}(u)\right|_{u=u_o}=0,
\end{equation}
thus
\begin{equation}
\label{CC4}
   u \approx  u_0 \simeq \frac{(j+1-k)c}{2r_{\rm A}}\,,
\end{equation}
where we have used the fact that $\alpha(iu)$ can be
regarded as almost constant for the small frequencies considered
here. For $j\le(k$ $\!-$ $\!1)$, $f_{jk}(u)$ is a monotonically
decreasing function, and relevant contributions to the frequency
integral can only come from regions, where
\begin{equation}
\label{CC5}
     u\le\frac{c}{2r_{\rm A}}\,,
\end{equation}
because for larger frequencies the exponentially decaying
factor becomes too small. Combining Eqs. (\ref{CC4}) and
(\ref{CC5}), it can be said that the relevant contributions
to the frequency integral come from regions, where $u$
$\!\lesssim$ $\!c/r_{\rm A}$. In these regions we have
\begin{equation}
\label{C6}
  |k_0R|,\ |kR| \sim \frac{uR}{c}
  \lesssim \frac{R}{r_{\rm A}}\,.
\end{equation}
This means that in the long-distance limit $r_{\rm A}$ $\!\gg$ $\!R$
the main contributions to the integral come from regions, where
$|k_0R|,\ |kR|\ll 1$. We may therefore expand the coefficients
$B^{M,N}_n$ in Eqs.~(\ref{E73}) and (\ref{E74})
in powers of $k_0R$, on exploiting useful relations in
Ref.~\cite{Abramowitz}, and retain only the leading terms:
\begin{equation}
\label{C7}
     B^M_n = o\left[(k_0R)^{2n+3}\right],
\end{equation}
\begin{equation}
\label{C8}
    B^N_n \simeq i\frac{(n+1)(2n+1)}{[(2n+1)!!]^2}
    \frac{\varepsilon(iu)-1} {\varepsilon(iu) n +n+1}(k_0R)^{2n+1}.
\end{equation}
In this way we find that the leading term in Eq.~(\ref{E75})
comes from the two terms containing $B^N_n$ with $n$ $\!=$ $\!1$.
Keeping only these terms,
using \cite{Abramowitz}
\begin{equation}
\label{C9}
   h^{(1)}_1(z)=-\left(\frac{1}{z}+\frac{i}{z^2}\right)e^{iz},
\end{equation}
and changing the integration variable according to
$u$ $\!\to$ $\!z$ $\!=$ $\!ur_{\rm A}/c$, we arrive at Eq.~(\ref{E77}).

\section{Derivation of Eq.~(\ref{E82})}
\label{AppE}

Provided that
\begin{equation}
\label{D1}
      n\gg \frac{|z|^2}{4}\,,
\end{equation}
%
the spherical Bessel and Hankel functions appearing in
Eqs.~(\ref{E73}) -- (\ref{E75})
can be approximated by
\cite{Abramowitz}
\begin{equation}
\label{jbign}
    j_n(z)\simeq \frac{z^n}{(2n+1)!!}\,,
\end{equation}
and
\begin{equation}
\label{ybign}
    h_n^{(1)}(z)\simeq -i \frac{(2n-1)!!}{z^{(n+1)}}\,,
\end{equation}
respectively.
As we know from Appendix \ref{AppD},
the main contribution to the frequency integral
in Eq.~(\ref{E75}) is from those values satisfying
the condition (\ref{CC5}).
Hence, the condition (\ref{D1}) becomes
\begin{equation}
\label{D2}
     n\gg 1,
\end{equation}
because $z$ $\!\sim$ $\!uR/c$ $\simeq\!ur_{\rm A}/c$ in the
short-distance limit.
Substituting in Eqs.~(\ref{E73}) -- (\ref{E75})
for $j_n(z)$ and $h_n^{(1)}(z)$ the expressions (\ref{jbign})
and (\ref{ybign}),
we derive after some algebra
\begin{equation}
\label{D5}
    B^M_n \simeq 0,
\end{equation}
\begin{eqnarray}
\label{D6}
    &&n(n+1)(2n+1)B^N_n \left[\frac{h^{(1)}_n(k_0r_{\rm A})}{k_0r_{\rm A}}\right]^2
\nonumber\\&&
    + (2n+1)B^N_n \left[\frac{\dif[r_{\rm A} h^{(1)}_n(k_0r_{\rm A})]}{k_0r_{\rm A}\dif r_{\rm A}}\right]^2
\nonumber\\&&
    \simeq
    -i\frac{1}{(k_0r_{\rm A})^3}
    \frac{\varepsilon(iu)-1}{\varepsilon(iu)+1}n(n+1)
   \left(\frac{R}{r_{\rm A}}\right)^{2n+1}
\nonumber\\&&\quad
    -i\frac{1}{(k_0r_{\rm A})^3}
    \frac{\varepsilon(iu)-1}{\varepsilon(iu)+1}
    \frac{(2n+1)^2}{4}
   \left(\frac{R}{r_{\rm A}}\right)^{2n+1}
\nonumber\\&&
    \simeq
    -2i\frac{1}{(k_0r_{\rm A})^3}
    \frac{\varepsilon(iu)-1}{\varepsilon(iu)+1}n(n+1)
   \left(\frac{R}{r_{\rm A}}\right)^{2n+1}.
\end{eqnarray}
Whereas in the long-distance limit we could
neglect all terms but the $n$ $\!=$ $\!1$ one
(Appendix \ref{AppD}), in
the short-distance limit, as can be seen
from Eq.~(\ref{D6}), the parameter $R/r_{\rm A}$ being very close to
one, we encounter the opposite extreme, where the main
contribution comes from those terms corresponding to high orders
$n$. The main contribution to the sum over $n$ in
Eq.~(\ref{E75}) comes from the peak at $n_1$ determined by
\begin{eqnarray}
\label{D7}
\lefteqn{
 \left.\frac{\dif}{\dif n}
 \left[n(n+1)\left(\frac{R}{r_{\rm A}}\right)^{2n+1}\right]\right|_{n=n_1}
}
\nonumber\\[1ex]&&
 \approx\left.\frac{\dif}{\dif n}\left(n^2e^{2n
 \left(\ln R-\ln r_{\rm A}\right)}\right)\right|_{n=n_1}
\nonumber\\[1ex]&&
  =2n_1\left(\frac{R}{r_{\rm A}}\right)^{2n_1}\left[1+n_1
  \left(\ln R-\ln r_{\rm A}\right)\right] =0,
\end{eqnarray}
from which we find
\begin{equation}
\label{D8}
   n_1=\frac{1}{\ln r_{\rm A}-\ln R}
   \simeq \frac{R}{\Delta r_{\rm A}}\,,
\end{equation}
because of
$\Delta r_{\rm A}/R$ $\!\ll$ $\!1$.
Since
the main contribution to the sum over $n$ in Eq.~(\ref{E75})
comes from values of \mbox{$n$ $\!\approx$ $\!n_1$ $\!\gg$ $\!1$},
where the approximate formulas (\ref{D5}) and (\ref{D6}) are valid
[cf. Eq. (\ref{D2})]. Therefore we introduce only a small error
if we extrapolate these formulas to the terms with small $n$.
Then the sum over $n$ is equal to the second derivative
with respect to $(R/r_{\rm A})^2$ of a geometric sum,
which can be performed in closed form to yield
Eq.~(\ref{E82}).


\end{document}